\documentclass[aps,english,prc,preprintnumbers,fleqn,showkeys,
superscriptaddress,floatfix]{revtex4}
\usepackage[T1]{fontenc}
\usepackage[latin9]{inputenc}
\usepackage{float}
\usepackage{bm}
\usepackage{graphicx}

\providecommand{\tabularnewline}{\\}

\usepackage{graphics}
\usepackage{epsf}
\usepackage{bm}
\usepackage{color}

\usepackage{babel}
\begin{document}
\preprint{INHA-NTG-05/2011}
%-----------------------------------------------
\title{Tensor properties of the nucleon}
%-----------------------------------------------

\author{Tim Ledwig}

\affiliation{Institut f\"ur Kernphysik, Universit\"at Mainz, D
-55099 Mainz, Germany} 

\author{Antonio Silva}

\affiliation{Faculdade de Engenharia da Universidade do Porto,
  R. Dr. Roberto Frias s/n, P-4200-465 Porto, Portugal} 

\affiliation{Centro de Fisica Computacional (CFC), P-3004-516 Coimbra, 
  Portugal} 

\author{Hyun-Chul Kim}

\email{hchkim@inha.ac.kr}

\affiliation{Department of Physics, Inha University, 
Incheon 402-751, Korea}

\date{March, 2011}
\begin{abstract}
We report in the present talk recent results of the tensor
properties of the nucleon within the framework of the chiral
quark-soliton model. The tensor and anomalous tensor magnetic form
factors are calculated for the momentum transfer up to 
$Q^{2}\leq1\,\mathrm{GeV}^{2}$ and at a renormalization scale of
$0.36\,\mathrm{GeV}^{2}$. The main results are summarized as follows:
the flavor tensor charges of the nucleon are yielded as $\delta
u=1.08$, $\delta d=-0.31$, $\delta s=-0.01$, while the up and down
anomalous tensor magnetic moments are evaluated as
$\kappa_{T}^{u}=3.56$ and $\kappa_{T}^{d}=1.83$, respectively. The
strange anomalous tensor magnetic moment turns out to be
$\kappa_{T}^{s}=0.2\sim -0.2$, compatible with zero. 
We discuss their physical implications, comparing them in particular
with those from the lattice QCD. 
\end{abstract}

%-----------------------------------------------
\maketitle

%-----------------------------------------------
\section{Introdcution}
%-----------------------------------------------
The transversity of the nucleon has been one of the most important
issues in hadronic physics well over decades. While the non-polarized
parton distribution functions (PDFs) and longitudinally polarized PDFs
have been extensively studied experimentally as well as theoretically,
the transversity of the nucleon suffers from experimental difficulties
to measure it. Only very recently, the tensor charges $\delta q$ were 
extracted~\cite{Anselmino:tensorcharge} from the
experiments~\cite{Belle:2006,HERMES:2005,COMPASS:2007}. 
The QCDSF and UKQCD lattice collaborations reported also the tensor
charges in the context of the spin structure of the
nucleon~\cite{Goeckeler:GPDlattice}.   

In this talk, we present recent results of the tensor and tensor
anomalous magnetic form factors within the framework of the
self-consistent SU(3) chiral quark-soliton model  
($\chi$QSM)~\cite{Ledwig:2010tu,Ledwig:2010zq}. 
The $\chi$QSM was known to describe successfully the mass splittings
of SU(3) baryons and static properties and form factors of the
nucleon~\cite{Christov:1995vm}. The model has certain merits: First it
does not have any adjustable parameters. The cut-off parameter is
fixed to the pion decay constant, the current quark mass is fitted in
such a way that it reproduces kaon properties. The constituent quark
mass, which is regarded as only one free parameter of the model, is
also fixed to 420 MeV with which the electric properties
of the nucleon are well reproduced. Secondly, it is a fully
relativistic field-theoretic model. We will see why it is essential to
describe the tensor properties of the nucleon.

Tensor charges of the nucleon were also investigated within this
framework~\cite{TensorSU2,TensorSU3}. However, the old version of the
SU(3) $\chi$QSM was hampered by the symmetry-nonconserving
quantization, which causes the breaking of gauge invariance.  
Thus, it is necessary to revisit the tensor charges of the nucleon
within the $\chi$QSM and we  present in this talk the
new results of the tensor charges and their form factors, and those of
the anomalous tensor magnetic form factors, closely following
Refs.~\cite{Ledwig:2010tu,Ledwig:2010zq}. We will also compare the
results of the present talk with the empirical data as well as those 
lattice calculation.  

%-----------------------------------------------
\section{Results and discussion} 
%-----------------------------------------------
The detailed formalism can be found in
Refs.~\cite{Ledwig:2010tu,Ledwig:2010zq}. We briefly report the
results of the tensor charges and anomalous tensor magnetic moments
and their form factors. The tensor charges are identical to the
axial-vector charges in the non-relativistic limit, i.e. the singlet
and triplet parts are $g_A^0=g_T^0=1$ and $g_A^3=g_T^3=5/3$,
respectively. In Table~\ref{tab:tensor_charge}, we summarize the
results of the tensor and axial-vector charges from the SU(2) and
SU(3) $\chi$QSM.
\begin{table}[h]
\caption{\label{tab:tensor_charge} Tensor charges for the singlet,
triplet, and octet components in comparison with the axial-vector
charges.}    
\begin{center}
\begin{tabular}{c|ccc|ccc}\hline
& $g_{T}^{0}$ &  $g_{T}^{3}$ & $g_{T}^{8}$ & $g_A^{0} $ & $g_A^3$ &
$g_A^8$ \tabularnewline 
\hline
SU(3) & $0.76$ & $1.40$ & $0.45$ & $0.37$  & $1.18$ & $0.36$
\tabularnewline 
SU(2) & $0.69$ & $1.45$ & $--$ & $0.36$  & $1.20$ & $--$
\tabularnewline 
\hline
\end{tabular}
\end{center}
\end{table}

The flavor tensor charges can be obtained by using the singlet,
triplet, and octet ones. In Table~\ref{tab:038-Tcharges}, the results
of the up, down, and strange tensor charges are listed. Note that the
tensor charges are scale-dependent. So, in order to compare the
present results with those of the lattice QCD, one needs to scale down
the lattice results to the scale of the $\chi$QSM, i.e. $\mu^2\approx
0.36\,\mathrm{GeV}^2$. The results of the  $\chi$QSM are comparable
with the lattice ones, as shown in Table~\ref{tab:038-Tcharges}. The
strange tensor charge turns out to be compatible with zero.  
\begin{table}[ht]
  \centering
  \caption{\label{tab:038-Tcharges}Tensor charges for each flavor in
    comparison with the axial-vector charges.}
  \label{tab:2}
  \begin{tabular}{c|cc|cc|cc} \hline
& $\Delta u$ & $\delta u$ & $\Delta d$ & $\delta d$ & $\Delta s$ &
$\delta s$   
\tabularnewline  \hline
$\chi$QSM & $0.84$ & $1.08$ & $-0.34$ & $-0.31$ & $-0.05$ & $-0.01$
\tabularnewline 
NRQM & $\frac{4}{3}$ & $\frac{4}{3}$ & $-\frac{1}{3}$ & $-\frac{1}{3}$ &
$0$ & $0$ \\
Lattice~\cite{Goeckeler:GPDlattice} & $$ & $1.05\pm0.16$ & $$ &
$-0.26\pm0.01$ & $--$ & $--$
\tabularnewline
\hline
  \end{tabular}
\end{table}

In Table~\ref{tab:COMP}, the numerical results of the anomalous tensor
magnetic moments are listed and compared with those of the lattice
QCD. While the up anomalous tensor magnetic moment is in good
agreement with the lattice but that for the down quark is smaller than
the lattice one. As a result, the ratio of the up and down anomalous
tensor magnetic moments turns out to be smaller than the lattice
result. 
\begin{table}[ht]
\caption{\label{tab:COMP} The results for the anomalous
  tensor magnetic moments $\kappa_T^q$ in comparison with the lattice 
  calculation. 
 }

\begin{tabular}{c|cc|c}\hline
 &Present work SU(3) & Present work SU(2)&  Lattice~\cite{Hagler:2008}
 \tabularnewline   
\hline
$\kappa_{T}^{u}$ & $3.56$&$3.72$  & $3.70$ \tabularnewline 
$\kappa_{T}^{d}$ & $1.83$&$1.83$  & $2.35$ \tabularnewline 
$\kappa_{T}^{s}$ & $0.2\sim-0.2$ & $ $
& \tabularnewline 
\hline
$\kappa_{T}^{u}/\kappa_{T}^{d}$ & $1.95$&$2.02$ & $1.58$ \tabularnewline \hline
\end{tabular}
\end{table}

\begin{figure}[ht]
\begin{center}
\includegraphics[scale=0.55]{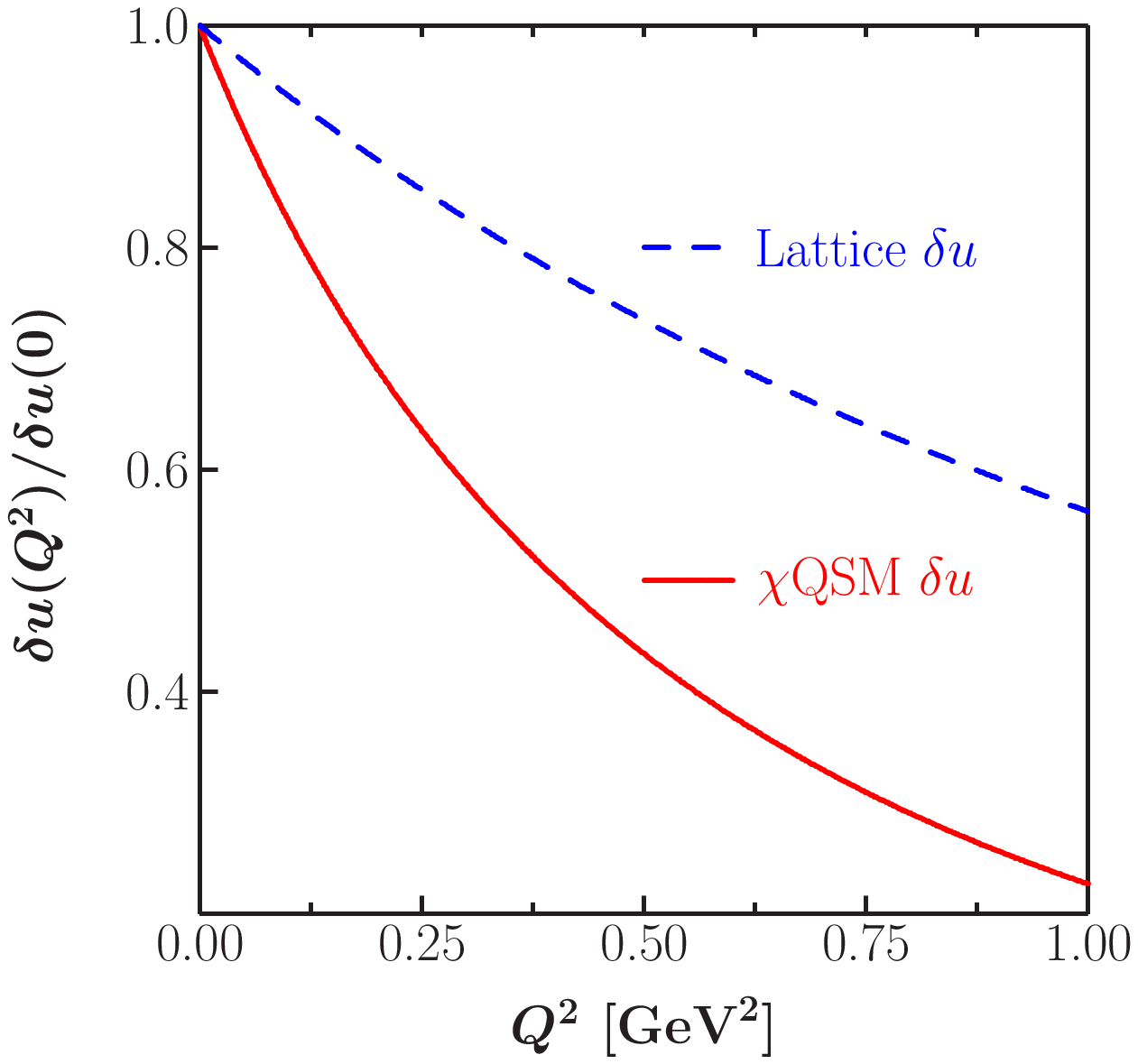}~~~
\includegraphics[scale=0.55]{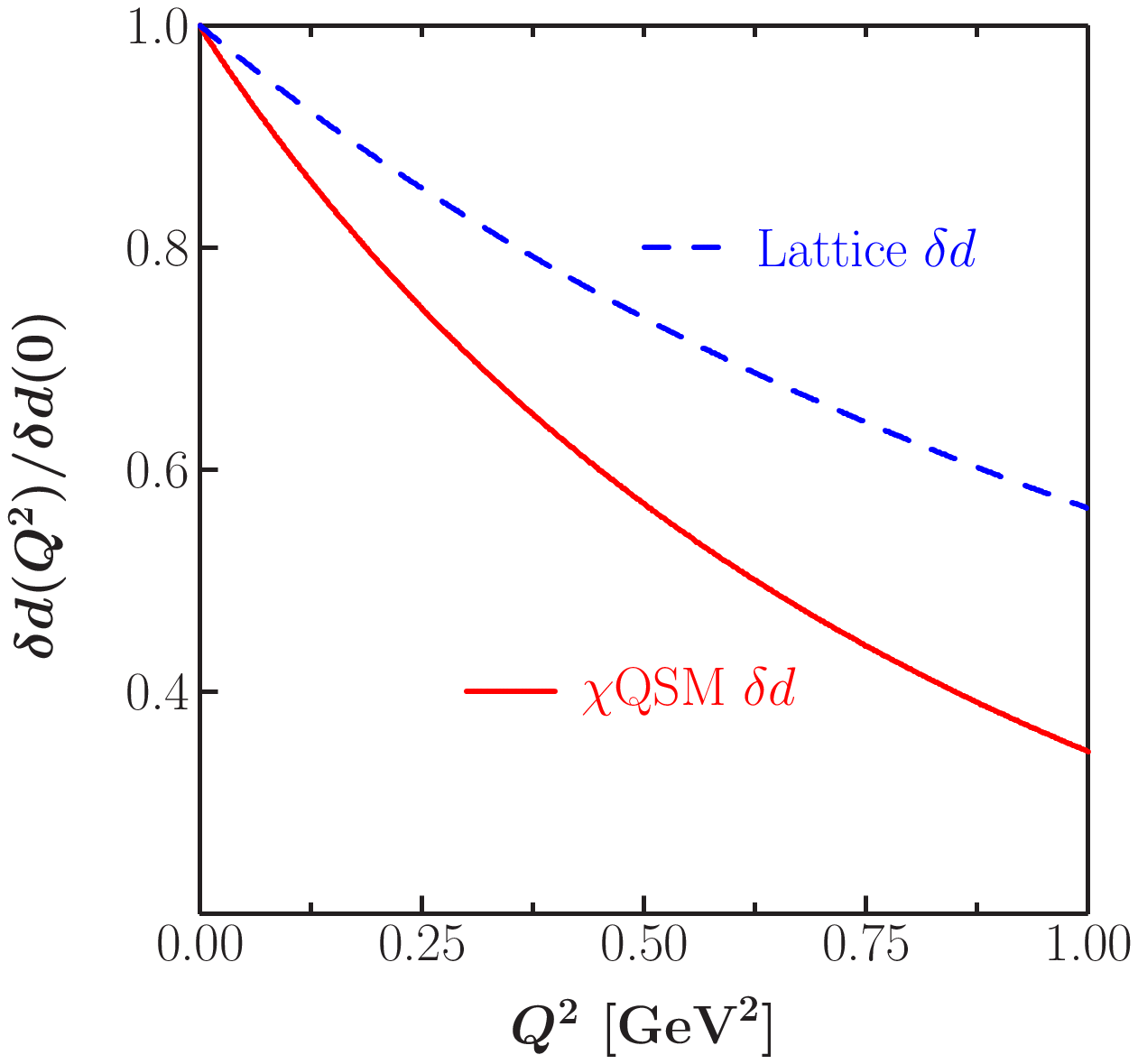}
\end{center}
\caption{\label{Fig:HT} flavor tensor form factors  
$\delta u(Q^{2})$, $\delta d(Q^{2})$ and $\delta s(Q^{2})$ for the
proton. We compare the present results of the
renormalization-independent scaled tensor form factors $\delta
u(Q^{2})/\delta u(0)$, $\delta d(Q^{2})/\delta d(0)$ with those of the
lattice QCD~\cite{Goeckeler:GPDlattice}. The red (solid) curves
designate the $\chi$QSM form factors of this work while the blue
(dashed) ones corrrespond to the factors from the lattice QCD
calculation.}    
\end{figure}

In Fig.~\ref{Fig:HT}, we draw the results of the scaled flavor tensor
form factors in comparison with those of the lattice calculation. 
Figure~\ref{Fig:HT} shows that the lattice results decrease in general
almost linearly as $Q^2$ increases. On the other hand, the 
present results fall off more rapidly. This is not a surprizing
result, because a similar behavior was also found in the
$\Delta(1232)$ electric quadrupole form factor as shown in
Ref.~\cite{CQSM:DeltaEQM}. These differences of the $Q^2$ dependence
may be due to the fact that the heavy pion mass was employed in the
lattice calculation. Moreover, this has been shown explicitly for the
nucleon isovector form factor $F_{1}^{p-n}(Q^{2})$ on the
lattice~\cite{Lattice:FuMd}.      

\begin{figure}[ht]

\begin{center}
\includegraphics[scale=0.55]{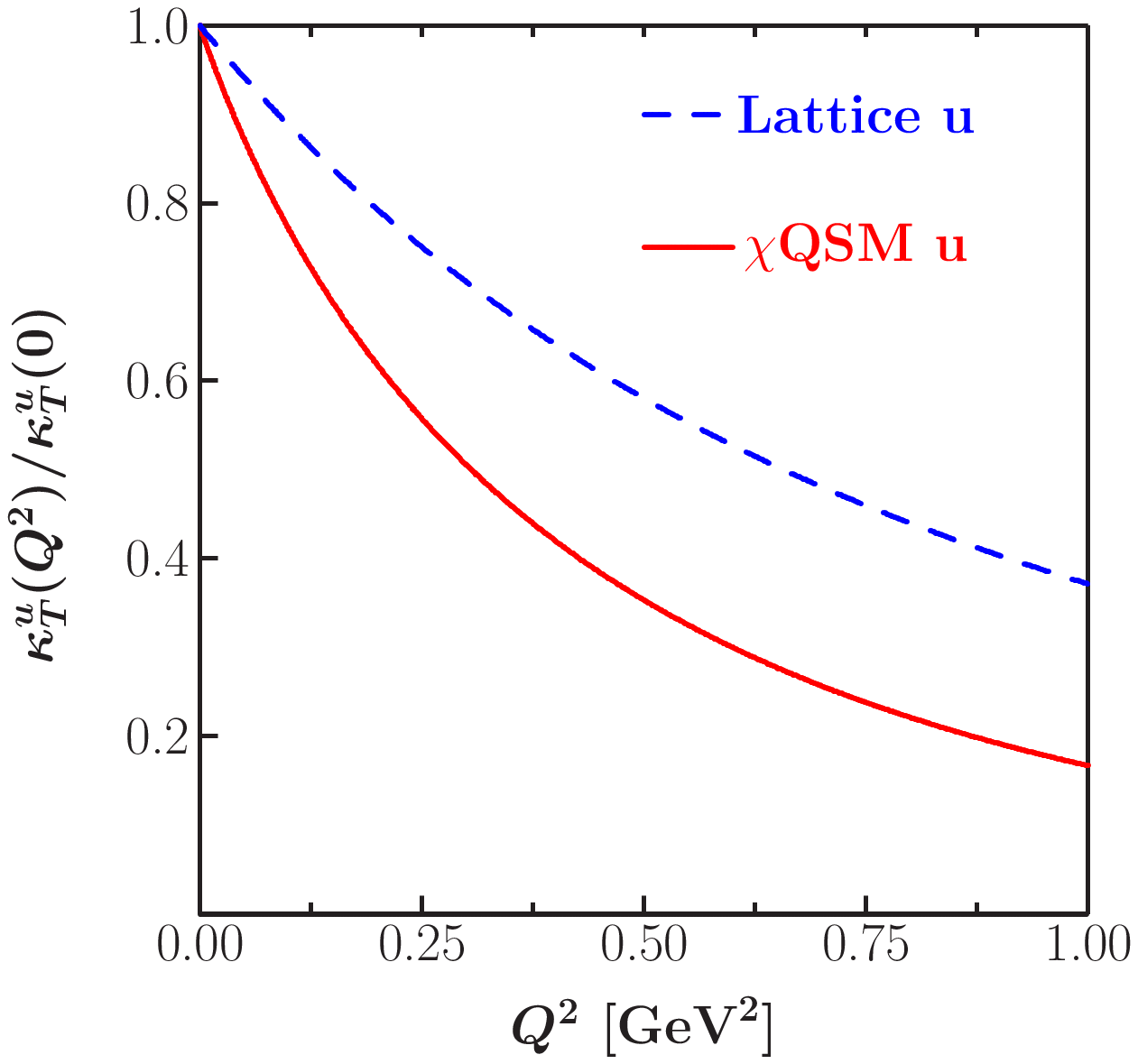}~~~
\includegraphics[scale=0.55]{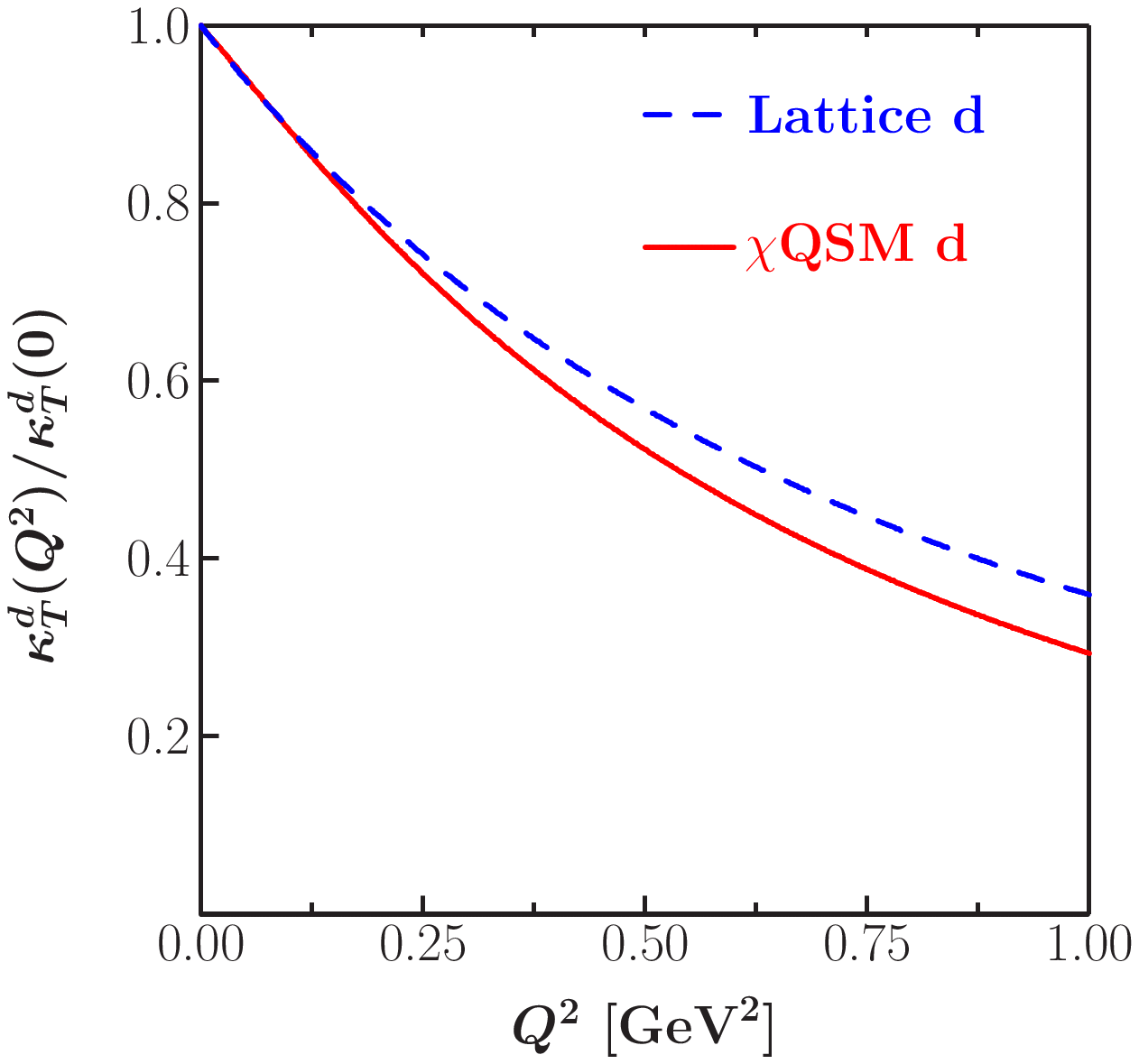}
\end{center}
\caption{\label{fig:Kapalat} The comparison of the
up and down anomalous tensor magnetic form factors with lattice
results. The solid curves draw the results of the present 
work with $M=420$ MeV, whereas the dashed ones depict the lattice
results~\cite{Hagler:2008}. In the left panel, the result of the up
anomalous tensor magnetic form factor is compared to that of the
lattice. The right panel is for the down form factors. The
lattice calculation was performed with $m_{\pi}=600$ MeV.}    
\end{figure}

Figure~\ref{fig:Kapalat} shows the comparison of the results of the
flavor anomalous tensor form factors with those from the lattice QCD.
Being similar to the case of the tensor form factors, the lattice
results decrease more slowly than those of the present work. However,
the down form factor is comparable to the lattice one. 
%-----------------------------------------------
\section{Summary and outlook}
%-----------------------------------------------
In the present talk, we have reported the recent results of the tensor
properties of the nucleon. The tensor charges and anomalous tensor
magnetic moments were calculated. The results are in good agreement
with those of the lattice calculation except for the down anomalous
tensor magnetic moment. The tensor form factors and anomalous tensor
form factors were also presented and compared with the lattice
results. 

The present results can be used to describe the transverse spin
structure of the nucleon. The corresponding investigation is under
way.  
%-----------------------------------------------
\section*{Acknowledgments}
%-----------------------------------------------
HChK is grateful to A.~Hosaka for the hospitality during the Baryon
2010. The present work is supported by Basic Science Research Program
through the National Research Foundation of Korea (NRF) funded by the
Ministry of Education, Science and Technology (grant number:
2010-0016265). 
%-----------------------------------------------

%--------------------------------------------------

\end{document}